\documentclass{article}
\usepackage{ijcai22}

\usepackage{times}
\usepackage{soul}
\usepackage{url}
\usepackage{color}
\usepackage[hidelinks]{hyperref}
\usepackage[utf8]{inputenc}
\usepackage[small]{caption}
\usepackage{graphicx}
\usepackage{amsmath}
\usepackage{amsfonts}
\usepackage{amsthm}
\usepackage{multirow}
\usepackage{booktabs}
\usepackage{algorithm}
\usepackage{algorithmic}
\title{AntPivot: Livestream Highlight Detection via Hierarchical Attention Mechanism}

\author{
    Anonymous
}

\author{
Yang Zhao$^{1}$\footnotemark[1]
\and
Xuan Lin$^{2}$\footnotemark[1]\and
Wenqiang Xu$^2$\and
Maozong Zheng$^2$\and
Zhengyong Liu$^2$\and
Zhou Zhao$^1$\footnotemark[2]
\affiliations
$^1$Zhejiang University \quad $^2$Ant Group\\
\emails
awalk@zju.edu.cn, daxuan.lx@antgroup.com \\
\{yugong.xwq, zhengmaozong.zmz, liuzhengyong.lzy\}@antgroup.com,
zhaozhou@zju.edu.cn
}

\begin{document}
\maketitle

\begin{abstract}
In recent days, streaming technology has greatly promoted the development in the field of livestream. Due to the excessive length of livestream records, it's quite essential to extract highlight segments with the aim of effective reproduction and redistribution. Although there are lots of approaches proven to be effective in the highlight detection for other modals, the challenges existing in livestream processing, such as the extreme durations, large topic shifts, much irrelevant information and so forth, heavily hamper the adaptation and compatibility of these methods. In this paper, we formulate a new task \textit{Livestream Highlight Detection}, discuss and analyze the difficulties listed above and propose a novel architecture \textit{AntPivot} to solve this problem. Concretely, we first encode the original data into multiple views and model their temporal relations to capture clues in a hierarchical attention mechanism. Afterwards, we try to convert the detection of highlight clips into the search for optimal decision sequences and use the fully integrated representations to predict the final results in a dynamic-programming mechanism. Furthermore, we construct a fully-annotated dataset \textit{AntHighlight} to instantiate this task and evaluate the performance of our model. The extensive experiments indicate the effectiveness and validity of our proposed method.
\end{abstract}
\renewcommand{\thefootnote}{\fnsymbol{footnote}}
\footnotetext[1]{Equal Contribution.}
\footnotetext[2]{Corresponding Author.}
\section{Introduction}
With the explosive growth of transmission speed and storage capacity on the Internet, an increasing amount of information with different levels of importance and usefulness is interwoven into various data flows. Meanwhile, there is also an irreversible tendency that people's available time is becoming more and more fragmented. As a result, users rarely have enough time or attention to separate the valuable information from the other useless part. For the sake of efficiency and convenience in the interaction, it's essential to extract key information from unprocessed data to help users get what they need with little effort. Under this requirement and circumstance, researchers try to design automatic algorithms to detect salient or highlight parts in different kinds of data.

Although plenty of approaches have emerged recently to process video, image or text in this field, the application and research in the area of livestream still suffer from lots of difficulties. First, contrary to other kinds of videos, livestreams are usually extremely long in duration, varying from dozens of minutes to several hours. Besides, a mass of noise and useless information, such as slips of the tongue, greetings, chit-chats and interactions with the audience, is also recorded as a part of livestreams, which harms the performance of methods to a large extent. Moreover, there always exist topic shifts and gaps in the expressions of streamers, resulting in low coherence and cohesion of corpus. Recently, some researchers also set about designing approaches to deal with the obstacles above. \cite{Fraser2020TemporalSO} choose to use audio transcripts and software logs to segment creative livestream, and \cite{Cho2021StreamHoverLT} propose \textit{StreamHover} to generate text summary for livestreams automatically. However, there is still a shortage of discussion on the scenarios where users need to watch the highlight clips retrieved from the original untrimmed record. A proper solution to this can not only help the redistribution and recreation of livestreams, but also bring a huge amount of benefits in the field of education and other industries by summarizing the online courses and significant conferences with the aid of machine intelligence.

To this end, we first formulate the task of \textit{Livestream Highlight Detection} as the segmentation and importance evaluation on the temporal dimension of livestreams. Considering there is no benchmark dataset available in this area, a bunch of livestream records in the domain of insurance and fortune are collected from the platform supported by AliPay to construct a new dataset called \textit{AntHighlight} to facilitate this task. To provide an elementary solution to accomplish the goal stated previously, we construct a novel architecture to extract and analyze the semantic information comprehensively and select highlight fragments from the untrimmed livestreams efficiently. Specifically, we first encode the raw data in different views and combine them into the inputs of our model. Afterwards, we utilize a novel hierarchical attention module, named \textit{Pivot Transformer}, to capture temporal dependencies and integrate representations from different semantic levels. Finally, a series of confidences and probabilities are calculated to determine the prediction results in a dynamic-programming manner. The detailed process and mechanism will be further described in the following sections.

In conclusion, the main contributions of this paper can be summarized in the following aspects:
\begin{itemize}
	\item We formulate the task of \textit{Livestream Highlight Detection} to explore the extraction of important livestream segments, which can be also regarded as a meaningful complement to the research in related areas.
	\item We propose a novel architecture named \textit{AntPivot} and introduce a newly-developed hierarchical attention module to address this problem in a dynamic-programming way with a proper budget of computation resource consumption.
	\item We collect a fully annotated dataset \textit{AntHighlight} from the livestream records in the domain of fortune and insurance and prove the feasibility and effectiveness of our proposed approach on this dataset.
\end{itemize}


\begin{figure*}[ht]
\centering
\includegraphics[width=\linewidth]{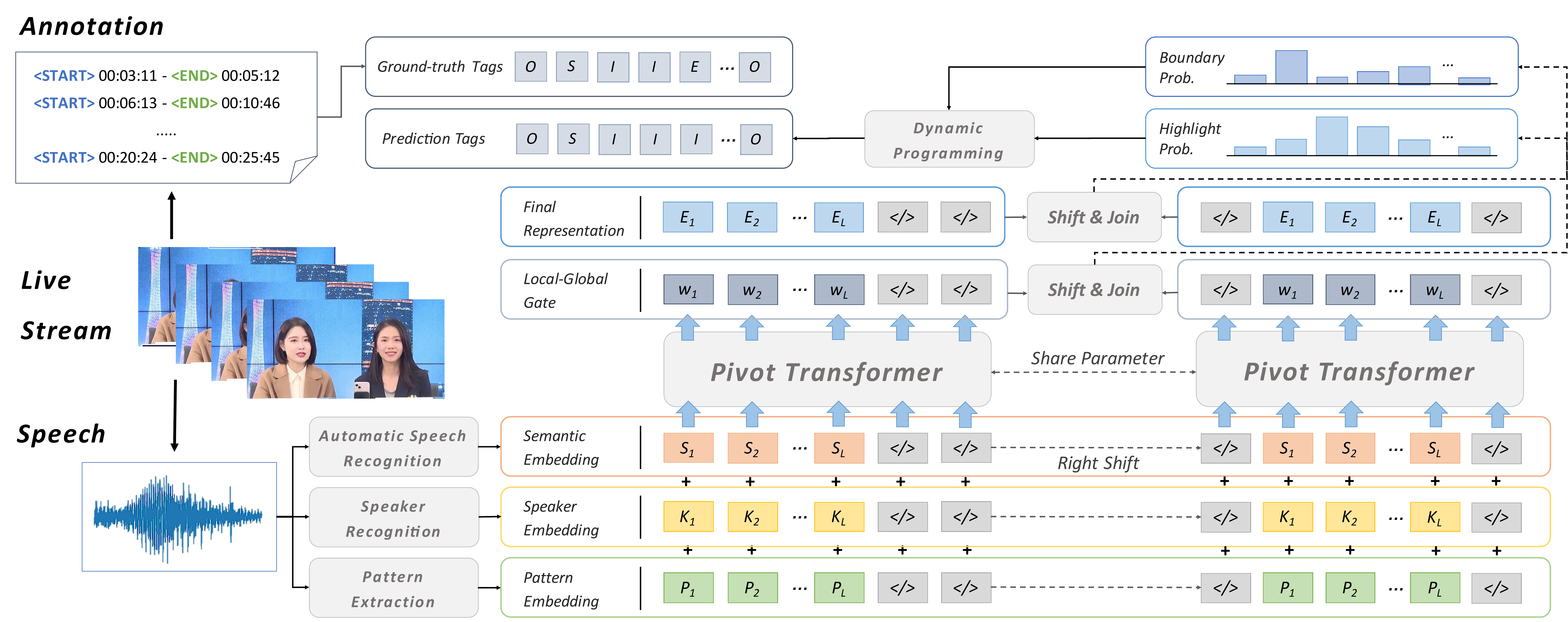}
\caption{Overall diagram of our proposed \textit{AntPivot} architecture. The $\scriptstyle </>$ annotation represents zero-pad in the sequence.}
\label{fig:overall}
\end{figure*}

\section{Related Work}

\paragraph{Text / Scene Segmentation} The task of text segmentation is to split documents or discourse into individual parts. Lexicon-based \cite{Hearst1997TextTS,Choi2000AdvancesID} and statistics-based approaches \cite{Utiyama2001ASM,Eisenstein2009HierarchicalTS} were initially applied to tackle this problem. Consequently, some efficient neural modules for sequence modeling, such as CRF \cite{Wang2018TowardFA}, PointerNetwork \cite{Li2018SegBotAG} and BERT \cite{Lukasik2020TextSB}, are also employed to boost a better generalization. Similarly, there also exist valuable discussions about the splitting of videos composed of complex scenes. Among them, early works \cite{Rasheed2003SceneDI,Chasanis2009SceneDI} try to utilize low-level features and carefully design heuristic methods. To explore supervised-learning strategies, some researchers construct a variety of new datasets based on documentaries \cite{Baraldi2015ADS}, short films \cite{Rotman2017OptimalSG}, long movies \cite{Rao2020ALA} etc. Different from text / scene segmentation, our proposed scenarios require the model to further analyze the interest and importance of every segment and accordingly filter out the useless and unimportant fragments.

\paragraph{Proposal Generation} Given an untrimmed video, the goal of action proposal generation is to ascertain a set of temporal boundaries with high probability or confidence to contain action instances. Current prevailing approaches can be mainly divided into two categories, namely anchor-based methods \cite{Gao2017TURNTT,Yang2021TemporalAP} and boundary-based ones \cite{Tan2021RelaxedTD,Su2021BSNCB}. The former choose candidates from hand-crafted proposal pre-definitions and the latter directly locate action boundaries.
Compared with this task, livestream highlight extraction mainly focuses on the comprehension and understanding of streamers' speech data, which have a higher semantic gap and lower redundancy than video data. Besides, the model in our scenario should not generate overlapping proposals, which is allowed and sometimes necessary in the task of action proposal generation.


\paragraph{Highlight Detection} Video highlight detection aims to compress a long video into the most informative series of frames or clips which can summarize the entire video with the highest density. In the early stage, \cite{Zhang2016VideoSW,Feng2018ExtractiveVS,Liu2019LearningHS} explore the temporal structure information to model dependencies between segments. Subsequently, \cite{Lal2019OnlineVS,Elfeki2019VideoSV,Huang2020ANK} try to further take the spatial positions and relations into consideration to measure importance score in a multi-dimensional mechanism. Although the highlight detection for livestream and video are quite similar inherently, the former task actually puts more emphasis on the structure and continuity of results, while the latter one tends to predict frame-level scores more individually.

\section{Method}

\subsection{Problem Definition} \label{chap:definition}
Given a long and unprocessed livestream record $\mathbf{R}$, the task of livestream highlight extraction aims to retrieve all proposals for highlight topics and discussions. To be specific, the livestream record can be annotated as $\mathbf{R} = (\mathbf{V}, \mathbf{A})$, where $\mathbf{V}$ and $\mathbf{A}$ represents visual and audio data respectively. And our goal is to construct a proper model to generate a series of proposals covering the most valuable parts of the entire livestream, which can be given as 
\begin{equation}
P=\{(s_1, e_1), (s_2, e_2), \dots ,(s_{p} , {e}_{{p}})\}
\end{equation}
where ${p}$ is the number of proposals and $({s}_i, {e}_i)$ is the start and end timestamp for $i$-th proposal satisfying 
$${s}_i < {e}_i < {s}_{i+1} < {e}_{i+1}, \forall i \in \{1, 2, \dots, {p}-1\}.$$

\subsection{Representation Inputs}
Figure \ref{fig:overall} demonstrates the overall architecture and calculation procedure of our proposed method. Before the next step of computing and reasoning, we need to project the original data into a continuous semantic space to learn some valuable patterns and structures to make decisions with the assistance of machine intelligence. To be specific, considering that the information contained in the changes of shots and scenes is less necessary than the one within the streamers' language expressions, we mainly focus on the processing of speech data in this paper. Having separated the audio data from livestream records, we first generate three kinds of embeddings to represent the data in multiple aspects, described as follows.

\paragraph{Semantic Embedding} Based on the same insight into the data with \cite{Fraser2020TemporalSO} that the linguistic expressions in livestreams serve as a vital and indispensable part in the analysis of information, we transform the speech data into the corresponding transcripts via an automatic speech recognition module.  Afterwards, a pretrained language model is employed to squeeze every sentence of transcripts into a single embedding, which can be annotated as $\mathbf{S} = \{\mathbf{s}_i\}_{i=1}^{L}$ where $L$ is the number of transcripts.

\paragraph{Speaker Embedding} Observing the records appearing in our scenario, there are lots of situations where multiple streamers jointly introduce some useful knowledge or good products in an interactive scheme, and their behaviors or expressions will be quite different according to their roles. Therefore, the identification of speakers may also help to localize important segments. In practice, we turn to the solution proposed in \cite{Wang2021AntVoiceNS} for effective speaker verification and then project the identification label into speaker embeddings, given by $\mathbf{K} = \{\mathbf{k}_i\}_{i=1}^{L}$.

\paragraph{Pattern Embedding} Besides, considering that most streamers tend to switch their mood, stress or pitch of voice when talking about something important or valuable to arouse the audience's attention and interest, we downsample the mel-frequency spectrum of speech pieces into a fixed length and concatenate them sequentially to seek out useful temporal patterns within every utterance, which is annotated as $\mathbf{P} = \{\mathbf{p}_i\}_{i=1}^{L}$. 

After obtaining all these embeddings, we compose them in an addition-based manner. Specifically, the final embeddings which are fed into our model can be given as $\mathbf{E} = \{\mathbf{e}_i\}_{i=1}^{L}$, where $\mathbf{e}_i = \mathbf{s}_i + \mathbf{p}_i + \mathbf{k}_i$ and all these embeddings are projected into the space of $\mathbb{R}^{d}$. \footnote{The dimension of representations keeps the same unless specified in the following sections.}
	

\begin{figure*}[ht]
\centering
\includegraphics[width=\linewidth]{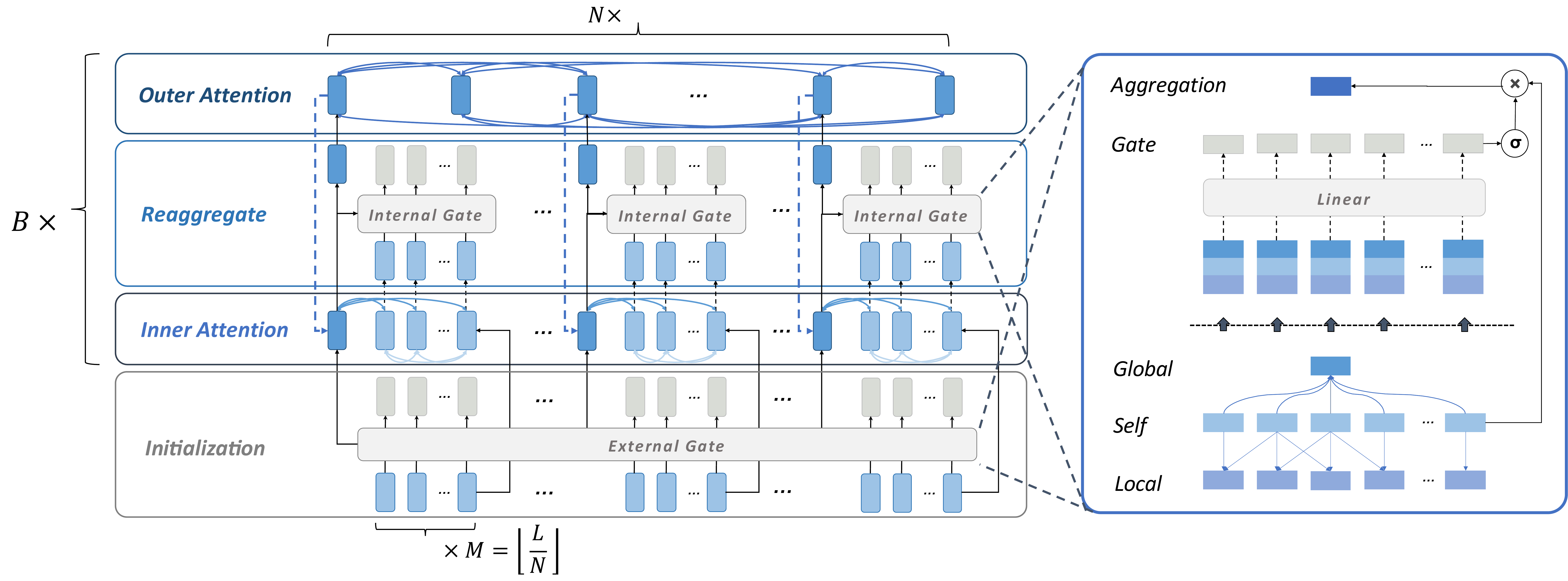}
\caption{Detailed structure of our proposed \textit{Pivot Transformer} and gating mechanism. The \textcolor[RGB]{53,91,183}{\textbf{blue dashed line}} represents dataflow between adjacent iterations.}
\label{fig:transformer}
\end{figure*}

\subsection{Pivot Transformer}
\label{chap:method}
Due to the large computation cost of global attention mechanism, it can be unbearable to utilize a vanilla transformer to deal with livestreams with a long duration under some constraints of devices. Additionally, there also exist massive topic shifts and irrelevant information, which makes the denoising and purification of information significant. In consideration of this, we devise a novel \textit{Pivot Transformer} to help model long-short term dependencies and filter out noise data within an appropriate and acceptable resource requirement. In this section, we will explain the working principle of this module concretely.

\paragraph{Initialization}
In the first stage, we employ a bi-directional gated recurrent unit proposed by \cite{Cho2014LearningPR} to generate the initial individual and global representations, given by
\begin{equation}
	\mathbf{g}, [\tilde{\mathbf{e}}_1, \cdots, \tilde{\mathbf{e}}_L] = \textit{Bi-GRU}([\mathbf{e}_1, \cdots, \mathbf{e}_L]),
\end{equation}
where $\mathbf{g}$ and $\tilde{\mathbf{e}}_i$ are the final state and output for the $i$-th step respectively.
After that, the input features will be divided into $N$ multiple consecutive blocks with the length of $M=\lfloor\frac{L}{N}\rfloor$, and the rearranged sequence of utterances can be ordered as 
\begin{equation}
\begin{aligned}
\mathbf{r}^0_{i,j} &= \mathbf{\tilde{e}}_{(i - 1)\times M+j}, & 1 \le i \le N, 1 \le j \le M,
\end{aligned}
\end{equation}
where $\mathbf{\mathbf{r}}^0_{i,j}$ represents the $j$-th element in the $i$-th block. And then, an \textit{External Gate} module (will be described below) is applied to generate a group of weights and dynamically aggregate elements into higher-level features which are called \textit{pivots} in this paper. The calculation can be formulated by 
\begin{equation}
\mathbf{w}^0_{i}, \mathbf{t}^0_i = \textit{ExternalGate}(\mathbf{r}^0_{i,1}, \cdots, \mathbf{r}^0_{i,M}), 
\end{equation} 
where $\mathbf{w}^0_{i} \in \mathbb{R}^{M}$ and $\mathbf{t}^0_i \in \mathbb{R}^{d}$ represent the gating weights and initial pivot of $i$-th block respectively. 

\paragraph{Update Mechanism}
Given the initialized pivots and elements, we try to exploit a hierarchical attention mechanism to incorporate context information efficiently. To be specific, in the $l$-th loop of interaction, a multi-head attention mechanism proposed in \cite{Vaswani2017AttentionIA} will be first adopted within every block to integrate local information, formulated as
\begin{equation}
[\tilde{\mathbf{t}}_i^l, {\mathbf{r}}_{i,1}^l, \cdots, {\mathbf{r}}_{i,M}^l] = \textit{MHA}([{\mathbf{t}}_i^{l-1}, {\mathbf{r}}_{i,1}^{l-1}, \cdots, {\mathbf{r}}_{i,M}^{l-1}])
\end{equation}
Subsequently, the pivot features will be reaggregated using an \textit{Internal Gate} module (will be described below), given by
\begin{equation}
\mathbf{w}_i^l, \hat{\mathbf{t}}_i^l = \textit{InternalGate}(\tilde{\mathbf{t}}_i^l, {\mathbf{r}}_{i,1}^l, \cdots, {\mathbf{r}}_{i,M}^l), 
\end{equation}
After this operation, the pivot representations can be treated as a reasonable and refined compression in a local range. To further capture the global context, another attention computation will be carried on the sequence of pivots, given as 
\begin{equation}
[\mathbf{t}_1^l, \cdots, \mathbf{t}^l_N] = \textit{MHA}([(\hat{\mathbf{t}}_1^l + \tilde{\mathbf{t}}_1^l), \cdots, (\hat{\mathbf{t}}_N^l + \tilde{\mathbf{t}}_N^l)])
\end{equation}
In the next step, the updated pivots will be fed into the $(l+1)$-th loop to pass information to the elements within every block and the overall procedure mentioned above will be repeated for a total of $B$ times. In this iterative calculation, the pivots actually act as a vital role to interact between local and global ranges. As for the final prediction, we first flatten the gating weights and updated representations as 
\begin{equation}
\begin{aligned}
	&\hat{w}^l_{(i-1) \times M + j} = (\mathbf{w}^l_{i})_j, &\hat{\mathbf{e}}_{(i-1) \times M + j} = \mathbf{r}^B_{i,j}, 
\end{aligned}
\end{equation}
and calculate the highlight confidences and boundary probabilities given by
\begin{equation}
\begin{aligned}
	&h_i = \frac{1}{B+1}\sum_{j=0}^B \sigma(\hat{w}^j_i), &b_i = \operatorname{MLP}(\hat{\mathbf{e}}_i),
\end{aligned}
\end{equation}
where $\operatorname{MLP}(\cdot)$ is a multi-layer perceptron module and $\sigma(\cdot)$ is the sigmoid function. 

It's worth mentioning that we copy the original sequence, shift the elements $M/2$ to the right and repeat the above operations in practice as shown in the right part of Figure \ref{fig:overall}, which is designed to keep the range of local receptive field at $3M/2$ to prevent the influence caused by absolute element positions. And all the final predictions will be shifted back and calculate the average results with the normal sequence.

\paragraph{Local-Global Gate} 
In this paragraph, we will introduce the gating mechanism used in the structure of pivot transformer, which is designed to estimate the importance of every utterance to predict the highlight score and ensure the essential information to be squeezed into the corresponding pivots. In the calculation, we consider the local context, global information and individual representation synthetically. As depicted in the framed region of Figure \ref{fig:transformer}, a group of global and local features will be generated and concatenated with the original sequence and then the gating weights will be produced via a multi-layer perceptron and be utilized to form the aggregation features, given as 
\begin{equation}
\begin{aligned}
	&w_i = \operatorname{MLP}([\mathbf{g};\mathbf{l}_i;\mathbf{r}_i]),
	&\mathbf{t} = \sum_i\frac{e^{w_i}}{\sum_j e^{w_j}}\mathbf{r}_i,
\end{aligned}
\end{equation}
where $[;]$ is the concatenation operator and $\mathbf{g}, \mathbf{l}_i, \mathbf{r}_i$ stand for the global, local and individual representations corresponding to the $i$-th element respectively.

According to the position of the gating unit in the structure, we further customize different schema for the \textit{external} and \textit{internal} ones. In the former ones, we utilize a 1D-convolution operator with a receptive field of $M/2$ to capture local information and take the final state of Bi-GRU as the global information. And for the latter, we directly treat the pivot features in the previous step as the local representations and calculate the global one by averaging all the elements in the sequence.
\paragraph{Complexity Analysis and Comparison}
Given the description of operations in this architecture, we analyze and compare the theoretical complexity of pivot transformer and the vanilla one in this part. By decomposing the original computation into a two-step hierarchical mechanism, the cost of $\mathcal{O}(L^2d)$ in the attention layer can be reduced to $\mathcal{O}(N \cdot M^2d + N^2d) = \mathcal{O}((\frac{L^2}{N} + N^2)d)$. The complexity achieves the optimum of $\mathcal{O}(L^{\frac{4}{3}}d)$ with $N = \Theta(L^{\frac{2}{3}})$, but in practice we set $N = M = \sqrt{L}$ for better efficiency in parallel computing.

\subsection{Training and Inference}
\paragraph{Training} 
As is shown in the architecture diagram and described in the sections above, the computation and prediction can be processed in an end-to-end manner. In order to enable our model to obtain knowledge from different aspects of supervision information, we design a multi-task loss function to train our model, including two parts listed as follows. The first one, named \textit{Boundary Loss}, is used to guide an accurate boundary recognition for the highlight clips, which can be formulated as
\begin{equation}
\mathcal{L}_{b} = -\sum_{i=1}^{L}(\bar{b}_i \operatorname{log}(b_i) + (1 - \bar{b}_i) \operatorname{log}(1-b_i)),
\end{equation}
However, it's still difficult for the model to address this problem without any extra knowledge, because there is still no adequate clues for the selection and filter of highlight parts. Therefore, we further apply the item of \textit{Highlight Loss} to help the model to discriminate highlight segments from others, given by
\begin{equation}
\mathcal{L}_{h} = -\sum_{i=1}^{L}(\bar{h}_i \operatorname{log}(h_i) + (1 - \bar{h}_i) \operatorname{log}(1-h_i)), 
\end{equation}
where $\bar{b}_i$ and $\bar{h}_i$ are the boundary and highlight indicators for the ground-truth annotation of $i$-th utterance respectively. 
Finally, the overall loss function in the training process can be composed in a weighted way like $\mathcal{L} = \mathcal{L}_{b} + \lambda \mathcal{L}_{h}$, where $\lambda$ is the hyper-parameter to balance these two parts.

\paragraph{Inference}
Given the highlight confidences $\{h_i\}^L_{i=1}$ and the boundary probabilities $\{b_i\}^L_{i=1}$, our goal is to ascertain an optimal prediction sequence to maximize the accumulative score on the final decision path. To make it clear, we first categorize all the possible states of the $i$-th element into four types. \romannumeral1) \textbf{start} / \romannumeral2) \textbf{end}: the $i$-th utterance is selected as the start / end boundary of some proposal; {\romannumeral3}) \textbf{in} / {\romannumeral4}) \textbf{out}: the $i$-th utterance locates in some valid proposal / isn't contained in any proposal.

And we use $f_{i,j}$ to represent the maximal score accumulated to the $i$-th element in the $j$-th state listed above. Therefore, the state transition equation can be designed as 
\begin{equation}
f_{i,j} = \left\{
\begin{aligned}
&\operatorname{max}(f_{i-1,2}, f_{i-1,4}) + b_i  &j &=1 \\
&\operatorname{max}(f_{i-1,1}, f_{i-1,3}) + b_i  &j &=2 \\
&\operatorname{max}(f_{i-1,1}, f_{i-1,3}) + h_i (1 - b_i)  &j &=3 \\
&\operatorname{max}(f_{i-1,2}, f_{i-1,4}) + (1 - h_i)(1 - b_i)  &j &=4
\end{aligned}
\right.
\end{equation}
where the initial states are set as $f_{1,1} = b_1$, $f_{1,2} = f_{1,3} = -\infty$ and $f_{1,4} = (1 - h_1)(1-b_1)$ because only the \textit{start} and \textit{out} state are legal for the first utterance.

Ultimately, we backtrack the optimal decision path from $f_{L, 3}$ or $f_{L, 4}$ to recover the entire sequence of choices and predictions. By doing so, we can examine all possible combinations of decisions holistically and obtain a stable and reliable result. Besides, from another point of view, the inference procedure can be also interpreted as a computation executed on a non-deterministic finite automaton.

\section{Experiment}

\subsection{Dataset} \label{chap:dataset}
\renewcommand\arraystretch{1.2}
\begin{table}[]
\begin{tabular}{lr}
\hline
\multicolumn{1}{|l}{Total number of livestream records}                     & \multicolumn{1}{r|}{3655}    \\
\multicolumn{1}{|l}{Total duration of records in hours}           & \multicolumn{1}{r|}{2861.55} \\
\multicolumn{1}{|l}{Total number of human annotators}           & \multicolumn{1}{r|}{10} \\
\multicolumn{1}{|l}{Average number of utterance in a record}     & \multicolumn{1}{r|}{298.58}  \\
\multicolumn{1}{|l}{Average number of utterance in a highlight}     & \multicolumn{1}{r|}{15.04}    \\
\multicolumn{1}{|l}{Average number of words in an utterance}     & \multicolumn{1}{r|}{24.63}   \\
\multicolumn{1}{|l}{Average number of highlight in a record}     & \multicolumn{1}{r|}{9.60}    \\
\multicolumn{1}{|l}{Average duration of highlight in seconds}    & \multicolumn{1}{r|}{211.87}  \\
\multicolumn{1}{|l}{Average percentage of utterance in a record} & \multicolumn{1}{r|}{91.63\%} \\
\multicolumn{1}{|l}{Average percentage of highlight in a record} & \multicolumn{1}{r|}{7.65\%}  \\
 \hline
\end{tabular}
\caption{Statistic information of \textit{AntHighlight} Dataset.}
\label{tab:dataset}
\end{table}

As a novel task, Livestream Highlight Detection lacks a proper and available dataset to serve as a benchmark. In light of this, we construct the \textit{AntHighlight} dataset by collecting a series of livestream records and annotating all the boundary timestamps for the highlight segments within them. 

Concretely, we first download a total of 3,655 livestream records, which mainly focus on the theme of funds and wealth, from the platform supported by Alipay. And all the expressions and interactions are in Chinese. For the sake of objectivity and rationality of annotations, we generally define highlights as the valuable segments with an integral topic and complete content, including the introductions to the recommended products, industry profiles, investment suggestions and so on. Besides, some meaningful interactions between users and streamers are regarded as highlights as well, including Q\&A, experience sharing, etc. Afterwards, 10 human annotators are employed to view the entire livestreams and annotate the ground-truth timestamp of start / end boundaries according to this definition. 

And for the purpose of training and evaluation, we further divide the annotated data into three subsets, including training, validation and test dataset with the size of 3055, 100 and 500 respectively. The detailed statistics and properties can be found in Table \ref{tab:dataset}.

\subsection{Metrics}
In order to evaluate the effectiveness of models objectively and automatically, we adopt two criteria widely used in the related fields, including \textbf{Average Precision} and \textbf{Boundary F1 Score}.

\paragraph{Average Precision} The task of livestream highlight detection aims to generate proposals to cover the target highlight parts tightly. Therefore, the quality of predictions is determined by the overlap with the ground-truth intervals. Following the conventional protocol in the area of action detection, we use Average Precision with tIoU thresholds \{0.5, 0.6, 0.7, 0.8, 0.9\} to measure the performance. 

\paragraph{Boundary F1 Score}
In addition to the IoU-based metrics, the evaluation of boundary classification should get concerned as well, because an accurate boundary prediction can not only boost the overall precision but also greatly reduce the expense of manual modification and revision to the final results. Considering the application scenarios in reality, we directly desert the intervals within 10 seconds and treat the predictions with a minimal difference of fewer than 5 seconds from ground-truth boundaries as correct ones. Take the evaluation of start timestamps as an example, the F1 score can be calculated as
\begin{equation}
\tilde{p} = \sum_{i=1}^{p}\mathbb{I}(\mathop{min}\limits_{j \in \{1, \cdots, \bar{p}\}}(|s_i - \bar{s}_j|) < 5),
\end{equation}
\begin{equation}
F_1 = \frac{2 \times Prec \times Rec}{Prec + Rec} = \frac{2\tilde{p}}{(\bar{p} + p)},
\end{equation}
where $\mathbb{I}(\cdot)$ is the indicator function, and $\{\bar{s}_i, \bar{e}_i\}_{i=1}^{\bar{p}}$ and $\{s_i, e_i\}_{i=1}^{p}$ stand for the ground-truth proposals and predicted ones.

\subsection{Comparison and Analysis}
In this section, we will conduct some experiments and ablation studies to compare and analyze the performances with different model structures or under various settings. All the details about implementations and hyper-parameter selections will be described in the supplementary materials.

\begin{table}[ht]
\centering
\begin{tabular}{c|ccccc|cc}
\toprule
\multirow{2}{*}{Inputs}      & \multicolumn{5}{c|}{Average Precision} & \multicolumn{2}{c}{F1} \\ \cline{2-8}
 & 0.5    & 0.6   & 0.7   & 0.8   & 0.9   & Start           & End           \\ \hline
S     & 67.2   & 58.2  & 50.9  & 43.1  & 34.6  & 42.2            & 45.1          \\
S+P   & 67.3   & 59.1  & 51.7  & 43.4  & 34.9  & 43.3            & 46.2          \\
S+K   & 67.8   & 59.6  & 52.2  & 43.6  & 35.1  & 43.2            & 46.8          \\
S+K+P & \textbf{68.9}   & \textbf{60.1}  & \textbf{52.6}  & \textbf{44.4}  & \textbf{36.0}  & \textbf{44.5} & \textbf{48.9}          \\ 
\bottomrule
\end{tabular}
\caption{Performance comparison on different combinations of modal inputs. The best results are given in \textbf{bold}.}
\label{tab:modal}
\end{table}

\begin{table}[ht]
\centering
\begin{tabular}{c|ccccc|cc}
\toprule
\multirow{2}{*}{Model}      & \multicolumn{5}{c|}{Average Precision} & \multicolumn{2}{c}{F1} \\ \cline{2-8}
            & 0.5   & 0.6   & 0.7   & 0.8   & 0.9   & Start      & End       \\ \hline
- Pivot      		& 66.7  & 58.2  & 50.9  & 42.5  & 34.3  & 43.1       & 47.2      \\
- Shift              & 67.9  & 59.7  & 52.3  & 43.8  & 35.4  & 41.9       & 45.8      \\
All & \textbf{68.9}   & \textbf{60.1}  & \textbf{52.6}  & \textbf{44.4}  & \textbf{36.0}  & \textbf{44.5} & \textbf{48.9} \\ 
\bottomrule
\end{tabular}
\caption{Performance comparison on different model settings.}
\label{tab:component}
\end{table}

\begin{table}[ht]
\centering
\begin{tabular}{c|ccccc|cc}
\toprule
\multirow{2}{*}{Model}      & \multicolumn{5}{c|}{Average Precision} & \multicolumn{2}{c}{F1} \\ \cline{2-8}
            & 0.5   & 0.6   & 0.7   & 0.8   & 0.9   & Start      & End       \\ \hline
GRU              & 66.6  & 57.9  & 51.0  & 42.9  & 34.6  & 43.2       & 47.8      \\
Trm.      & 67.9  & 59.9  & \textbf{52.8}  & \textbf{44.5}  & \textbf{36.0}  & 44.3       & 48.5      \\
Pvt.      & \textbf{68.9}  & \textbf{60.1}  & 52.6  & 44.4  & \textbf{36.0}  & \textbf{44.5}       & \textbf{48.9}      \\
\bottomrule
\end{tabular}
\caption{Performance comparison on different choices of modules.}
\label{tab:structure}
\end{table}

\begin{table}[ht]
\centering
\begin{tabular}{c|ccccc|cc}
\toprule
\multirow{2}{*}{Strategy}      & \multicolumn{5}{c|}{Average Precision} & \multicolumn{2}{c}{F1} \\ \cline{2-8}
            & 0.5   & 0.6   & 0.7   & 0.8   & 0.9   & Start      & End      \\ \hline
Simple      		& 67.8  & 58.0  & 50.6  & 40.6  & 30.9  & 36.8       & 42.2     \\
Greedy              & 63.5  & 56.1  & 49.2  & 42.4  & 35.1  & 42.8       & 44.7     \\
Ours & \textbf{68.9}   & \textbf{60.1}  & \textbf{52.6}  & \textbf{44.4}  & \textbf{36.0}  & \textbf{44.5} & \textbf{48.9} \\ 
\bottomrule
\end{tabular}
\caption{Performance comparison on different inference strategies.}
\label{tab:strategy}
\end{table}

\paragraph{Analysis on Modal Inputs} Table \ref{tab:modal} demonstrates the performance difference between multiple settings of modal input combinations. The capital letters ``S'' / ``K'' / ``P'' represent the usage of semantic / speaker / pattern embeddings, respectively. We can easily find that our model can achieve the best performance with all three kinds of embeddings, which shows the effectiveness of our proposed data processing schema. In addition, it can be observed from the table that the existence of speaker embedding has a larger impact than that of pattern embedding, which indicates that speaker identification is more necessary in the highlight detection.

\paragraph{Analysis on Model Structure} 
To verify the function and utility of our proposed \textit{Pivot Transformer}, we carry out some experiments on this in depth. Table \ref{tab:component} displays the evaluation results of replacing this component with other alternatives, where ``GRU'', ``Trm.'' and ``Pvt.'' are short for the bi-directional gated recurrent unit, vanilla transformer and pivot transformer, respectively. From this table, we can find that the pivot transformer achieves a comparable result with the vanilla one, and both of them surpass the GRU module. This observation suggests that the long-term memory indeed makes a difference in this task, which can be easily captured and maintained by Transformer-based architecture. Moreover, the distant information can be further denoised and compressed by our proposed mechanism, resulting in better complexity with little loss of accuracy. 

Apart from this, we also investigate the inner structure of this architecture. As shown in Table \ref{tab:structure}, we remove the shifting process and pivot mechanism (i.e. only the attention computation inside each block is conducted) to verify their effects. Without the interaction between pivot elements, the global information cannot get transmitted and utilized in the calculation, thus hindering the model from understanding the entire content comprehensively. Besides, there exists an apparent margin on the boundary F1 score in the absence of shifting procedure, which infers this operation can alleviate the impact brought by the absolute position of elements and enhance the discriminative ability in the local range.

\paragraph{Analysis on Prediction Strategy} 
To assess the effect of different prediction approaches in the inference stage, we further develop two other strategies to compare with the dynamic-programming method. \uppercase\expandafter{\romannumeral1}) \textbf{Simple}: we directly pick out all boundary candidates with the constraint of threshold $t_b$ and select all proposals with an average highlight score greater than threshold $t_h$. \uppercase\expandafter{\romannumeral2}) \textbf{Greedy}: We convert this task into a multi-class problem in this setting and make pairs between positions predicted as ``start'' and ``end'' categories. From Table \ref{tab:strategy}, our proposed strategy behaves best among them and the \textit{Simple} one is much inferior to the others. The reason can be inferred intuitively that the \textit{Simple} strategy disrupts the order of precedence, thus impeding the model from distinguishing the boundary type (i.e. start or end), and the assignment of threshold restricts the generalization of this setting. As for the \textit{Greedy} approach, it actually ignores the influence of relative differences in confidences and probabilities, resulting in a coarse and inaccurate result.


%

\section{Conclusion}
In this paper, we propose a novel livestream highlight detection task and collect a fully annotated dataset \textit{AntHighlight} to serve as a benchmark. To deal with this problem, we develop a hierarchical attention mechanism and gating strategy to integrate information efficiently and generate predictions in a dynamic-programming manner. The comprehensive experiments demonstrate the feasibility and effectiveness of our proposed method.

\appendix

\bibliographystyle{named}
\bibliography{ijcai22}

\begin{thebibliography}{}

\bibitem[\protect\citeauthoryear{Baraldi \bgroup \em et al.\egroup
  }{2015}]{Baraldi2015ADS}
Lorenzo Baraldi, Costantino Grana, and Rita Cucchiara.
\newblock A deep siamese network for scene detection in broadcast videos.
\newblock {\em Proceedings of the 23rd ACM international conference on
  Multimedia}, 2015.

\bibitem[\protect\citeauthoryear{Chasanis \bgroup \em et al.\egroup
  }{2009}]{Chasanis2009SceneDI}
Vasileios Chasanis, Aristidis Likas, and Nikolas~P. Galatsanos.
\newblock Scene detection in videos using shot clustering and sequence
  alignment.
\newblock {\em IEEE Transactions on Multimedia}, 11:89--100, 2009.

\bibitem[\protect\citeauthoryear{Cho \bgroup \em et al.\egroup
  }{2014}]{Cho2014LearningPR}
Kyunghyun Cho, Bart van Merrienboer, Çaglar G{\"u}lçehre, Dzmitry Bahdanau,
  Fethi Bougares, Holger Schwenk, and Yoshua Bengio.
\newblock Learning phrase representations using rnn encoder–decoder for
  statistical machine translation.
\newblock In {\em EMNLP}, 2014.

\bibitem[\protect\citeauthoryear{Cho \bgroup \em et al.\egroup
  }{2021}]{Cho2021StreamHoverLT}
Sangwoo Cho, Franck Dernoncourt, Tim Ganter, Trung Bui, Nedim Lipka, Walter
  Chang, Hailin Jin, Jonathan Brandt, Hassan Foroosh, and Fei Liu.
\newblock Streamhover: Livestream transcript summarization and annotation.
\newblock {\em ArXiv}, abs/2109.05160, 2021.

\bibitem[\protect\citeauthoryear{Choi}{2000}]{Choi2000AdvancesID}
Freddy Y.~Y. Choi.
\newblock Advances in domain independent linear text segmentation.
\newblock In {\em ANLP}, 2000.

\bibitem[\protect\citeauthoryear{Eisenstein}{2009}]{Eisenstein2009HierarchicalTS}
Jacob Eisenstein.
\newblock Hierarchical text segmentation from multi-scale lexical cohesion.
\newblock In {\em NAACL}, 2009.

\bibitem[\protect\citeauthoryear{Elfeki and Borji}{2019}]{Elfeki2019VideoSV}
Mohamed Elfeki and Ali Borji.
\newblock Video summarization via actionness ranking.
\newblock {\em 2019 IEEE Winter Conference on Applications of Computer Vision
  (WACV)}, pages 754--763, 2019.

\bibitem[\protect\citeauthoryear{Feng \bgroup \em et al.\egroup
  }{2018}]{Feng2018ExtractiveVS}
Litong Feng, Ziyin Li, Zhanghui Kuang, and Wayne Zhang.
\newblock Extractive video summarizer with memory augmented neural networks.
\newblock {\em Proceedings of the 26th ACM international conference on
  Multimedia}, 2018.

\bibitem[\protect\citeauthoryear{Fraser \bgroup \em et al.\egroup
  }{2020}]{Fraser2020TemporalSO}
C.~Ailie Fraser, Joy Kim, Hijung Shin, Joel Brandt, and Mira Dontcheva.
\newblock Temporal segmentation of creative live streams.
\newblock {\em Proceedings of the 2020 CHI Conference on Human Factors in
  Computing Systems}, 2020.

\bibitem[\protect\citeauthoryear{Gao \bgroup \em et al.\egroup
  }{2017}]{Gao2017TURNTT}
J.~Gao, Zhenheng Yang, Chen Sun, Kan Chen, and Ramakant Nevatia.
\newblock Turn tap: Temporal unit regression network for temporal action
  proposals.
\newblock {\em 2017 IEEE International Conference on Computer Vision (ICCV)},
  pages 3648--3656, 2017.

\bibitem[\protect\citeauthoryear{Hearst}{1997}]{Hearst1997TextTS}
Marti~A. Hearst.
\newblock Text tiling: Segmenting text into multi-paragraph subtopic passages.
\newblock {\em Comput. Linguistics}, 23:33--64, 1997.

\bibitem[\protect\citeauthoryear{Huang and Wang}{2020}]{Huang2020ANK}
Cheng Huang and Hongmei Wang.
\newblock A novel key-frames selection framework for comprehensive video
  summarization.
\newblock {\em IEEE Transactions on Circuits and Systems for Video Technology},
  30:577--589, 2020.

\bibitem[\protect\citeauthoryear{Lal \bgroup \em et al.\egroup
  }{2019}]{Lal2019OnlineVS}
Shamit Lal, Shivam Duggal, and Indu Sreedevi.
\newblock Online video summarization: Predicting future to better summarize
  present.
\newblock {\em 2019 IEEE Winter Conference on Applications of Computer Vision
  (WACV)}, pages 471--480, 2019.

\bibitem[\protect\citeauthoryear{Li \bgroup \em et al.\egroup
  }{2018}]{Li2018SegBotAG}
J.~Li, Aixin Sun, and Shafiq~R. Joty.
\newblock Segbot: A generic neural text segmentation model with pointer
  network.
\newblock In {\em IJCAI}, 2018.

\bibitem[\protect\citeauthoryear{Liu \bgroup \em et al.\egroup
  }{2019}]{Liu2019LearningHS}
Yen-Ting Liu, Yu-Jhe Li, Fu-En Yang, Shang-Fu Chen, and Y.~Wang.
\newblock Learning hierarchical self-attention for video summarization.
\newblock {\em 2019 IEEE International Conference on Image Processing (ICIP)},
  pages 3377--3381, 2019.

\bibitem[\protect\citeauthoryear{Lukasik \bgroup \em et al.\egroup
  }{2020}]{Lukasik2020TextSB}
Michal Lukasik, Boris Dadachev, Gonccalo Simoes, and Kishore Papineni.
\newblock Text segmentation by cross segment attention.
\newblock {\em ArXiv}, abs/2004.14535, 2020.

\bibitem[\protect\citeauthoryear{Rao \bgroup \em et al.\egroup
  }{2020}]{Rao2020ALA}
Anyi Rao, Linning Xu, Yu~Xiong, Guodong Xu, Qingqiu Huang, Bolei Zhou, and
  Dahua Lin.
\newblock A local-to-global approach to multi-modal movie scene segmentation.
\newblock {\em 2020 IEEE/CVF Conference on Computer Vision and Pattern
  Recognition (CVPR)}, pages 10143--10152, 2020.

\bibitem[\protect\citeauthoryear{Rasheed and Shah}{2003}]{Rasheed2003SceneDI}
Zeeshan Rasheed and Mubarak Shah.
\newblock Scene detection in hollywood movies and tv shows.
\newblock {\em 2003 IEEE Computer Society Conference on Computer Vision and
  Pattern Recognition, 2003. Proceedings.}, 2:II--343, 2003.

\bibitem[\protect\citeauthoryear{Rotman \bgroup \em et al.\egroup
  }{2017}]{Rotman2017OptimalSG}
Daniel Rotman, Dror Porat, and Gal Ashour.
\newblock Optimal sequential grouping for robust video scene detection using
  multiple modalities.
\newblock {\em Int. J. Semantic Comput.}, 11:193--208, 2017.

\bibitem[\protect\citeauthoryear{Su \bgroup \em et al.\egroup
  }{2021}]{Su2021BSNCB}
Haisheng Su, Weihao Gan, Wei Wu, Junjie Yan, and Y.~Qiao.
\newblock Bsn++: Complementary boundary regressor with scale-balanced relation
  modeling for temporal action proposal generation.
\newblock In {\em AAAI}, 2021.

\bibitem[\protect\citeauthoryear{Tan \bgroup \em et al.\egroup
  }{2021}]{Tan2021RelaxedTD}
Jing Tan, Jiaqi Tang, Limin Wang, and Gangshan Wu.
\newblock Relaxed transformer decoders for direct action proposal generation.
\newblock {\em ArXiv}, abs/2102.01894, 2021.

\bibitem[\protect\citeauthoryear{Utiyama and Isahara}{2001}]{Utiyama2001ASM}
Masao Utiyama and Hitoshi Isahara.
\newblock A statistical model for domain-independent text segmentation.
\newblock In {\em ACL}, 2001.

\bibitem[\protect\citeauthoryear{Vaswani \bgroup \em et al.\egroup
  }{2017}]{Vaswani2017AttentionIA}
Ashish Vaswani, Noam~M. Shazeer, Niki Parmar, Jakob Uszkoreit, Llion Jones,
  Aidan~N. Gomez, Lukasz Kaiser, and Illia Polosukhin.
\newblock Attention is all you need.
\newblock {\em ArXiv}, abs/1706.03762, 2017.

\bibitem[\protect\citeauthoryear{Wang \bgroup \em et al.\egroup
  }{2018}]{Wang2018TowardFA}
Yizhong Wang, Sujian Li, and Jingfeng Yang.
\newblock Toward fast and accurate neural discourse segmentation.
\newblock In {\em EMNLP}, 2018.

\bibitem[\protect\citeauthoryear{Wang \bgroup \em et al.\egroup
  }{2021}]{Wang2021AntVoiceNS}
Zhiming Wang, Furong Xu, Kaisheng Yao, Yuan Cheng, Tao Xiong, and Huijia Zhu.
\newblock Antvoice neural speaker embedding system for ffsvc 2020.
\newblock {\em Interspeech 2021}, 2021.

\bibitem[\protect\citeauthoryear{Yang \bgroup \em et al.\egroup
  }{2021}]{Yang2021TemporalAP}
Haosen Yang, Wenhao Wu, Lining Wang, Sheng Jin, Boyang Xia, Hongxun Yao, and
  Hujie Huang.
\newblock Temporal action proposal generation with background constraint.
\newblock 2021.

\bibitem[\protect\citeauthoryear{Zhang \bgroup \em et al.\egroup
  }{2016}]{Zhang2016VideoSW}
Ke~Zhang, Wei-Lun Chao, Fei Sha, and Kristen Grauman.
\newblock Video summarization with long short-term memory.
\newblock In {\em ECCV}, 2016.

\end{thebibliography}


\begin{thebibliography}{}

\bibitem[\protect\citeauthoryear{Loshchilov and
  Hutter}{2016}]{Loshchilov2016SGDRSG}
Ilya Loshchilov and Frank Hutter.
\newblock Sgdr: Stochastic gradient descent with restarts.
\newblock {\em ArXiv}, abs/1608.03983, 2016.

\bibitem[\protect\citeauthoryear{Loshchilov and
  Hutter}{2019}]{Loshchilov2019DecoupledWD}
Ilya Loshchilov and Frank Hutter.
\newblock Decoupled weight decay regularization.
\newblock In {\em ICLR}, 2019.

\bibitem[\protect\citeauthoryear{Reimers and
  Gurevych}{2020}]{reimers-2020-multilingual-sentence-bert}
Nils Reimers and Iryna Gurevych.
\newblock Making monolingual sentence embeddings multilingual using knowledge
  distillation.
\newblock In {\em Proceedings of the 2020 Conference on Empirical Methods in
  Natural Language Processing}. Association for Computational Linguistics, 11
  2020.

\bibitem[\protect\citeauthoryear{Vaswani \bgroup \em et al.\egroup
  }{2017}]{Vaswani2017AttentionIA}
Ashish Vaswani, Noam~M. Shazeer, Niki Parmar, Jakob Uszkoreit, Llion Jones,
  Aidan~N. Gomez, Lukasz Kaiser, and Illia Polosukhin.
\newblock Attention is all you need.
\newblock {\em ArXiv}, abs/1706.03762, 2017.

\bibitem[\protect\citeauthoryear{Wang \bgroup \em et al.\egroup
  }{2021}]{Wang2021AntVoiceNS}
Zhiming Wang, Furong Xu, Kaisheng Yao, Yuan Cheng, Tao Xiong, and Huijia Zhu.
\newblock Antvoice neural speaker embedding system for ffsvc 2020.
\newblock {\em Interspeech 2021}, 2021.

\bibitem[\protect\citeauthoryear{Zhang \bgroup \em et al.\egroup
  }{2020}]{Zhang2020StreamingCM}
Shiliang Zhang, Zhifu Gao, Haoneng Luo, Ming Lei, Jie Gao, Zhijie Yan, and Lei
  Xie.
\newblock Streaming chunk-aware multihead attention for online end-to-end
  speech recognition.
\newblock {\em ArXiv}, abs/2006.01712, 2020.

\end{thebibliography}

\end{document}


\maketitle

\section{Implementation Details}
\subsection{Data Processing}
\paragraph{Semantic Embedding} Any off-the-shelf automatic speech recognition module and language model pretrained on Chinese corpus can be used to produce semantic embeddings. In our experiments, we employ the \textit{LC-SAN-M} proposed by \cite{Zhang2020StreamingCM} to generate transcripts, which is pretrained on a 20000-hour Mandarin ASR task and finetuned in a 60-hour Mandarin corpus, and we extract the representations corresponding to the [CLS] token predicted by \textit{Sentence-BERT} introduced in \cite{reimers-2020-multilingual-sentence-bert} as the sentence embeddings. At the last step, we project the initial 768-d features into 256-d ones using a multi-layer perceptron.

\paragraph{Speaker Embedding} To produce speaker embeddings, we first adopt the approach proposed by \cite{Wang2021AntVoiceNS} to generate speaker labels for every utterance. And then, we use a 256-d lookup-table to project identification labels into continuous embeddings.

\paragraph{Pattern Embedding} In this part, we first generate the 128-d logarithm mel-filterbanks from every utterance and downsample them into a fixed length of 8, resulting in the representations with the size of $(L, 8, 128)$. Afterwards, we concatenate them into single vectors with the length of 1024 and project them into the 256-d subspace via a multi-layer perceptron.

\subsection{Model Setting}
Considering the maximal sequence length reaches about 900, we set $N = \sqrt{L} = 30$ as mentioned in the Chapter 3.3. This setting is also available for the situation where $L > 900$, and the practical complexity will increase accordingly. The dimension $d$ used in our model is set as 256. For all transformer-based architecture, the number of heads is 4 and sandwich layernorm mechanism (i.e. both pre-LN and post-LN are utilized) is adopted. In the pivot transformer, $B$ is set as 3 and the number of attention layer stacked in every stage is 2. As for the vanilla transformer and GRU in the experiment part, the number of layers is set as 6 to keep consistency. 

\subsection{Optimization and Inference}
All the experiments are conducted on one piece of Tesla P100. In the training procedure, we employ AdamW optimizer proposed by \cite{Loshchilov2019DecoupledWD} with warmup strategy \cite{Vaswani2017AttentionIA} and cosine annealing learning \cite{Loshchilov2016SGDRSG}. The maximal learning rate of is set as 3e-4 and the weight decay is fixed at 5e-5. To prevent overfitting, a dropout strategy with $p=0.4$ is applied in the structure. The training will last for 20 epochs and we select the checkpoint with the best performance on the validation dataset. And in the inference stage, all the scores will be pre-processed via a min-max normalization. The thresholds in \textit{Simple} strategy are set as $t_b = 0.25$ and $t_h = 0.7$.

\begin{figure}
\includegraphics[width=\linewidth]{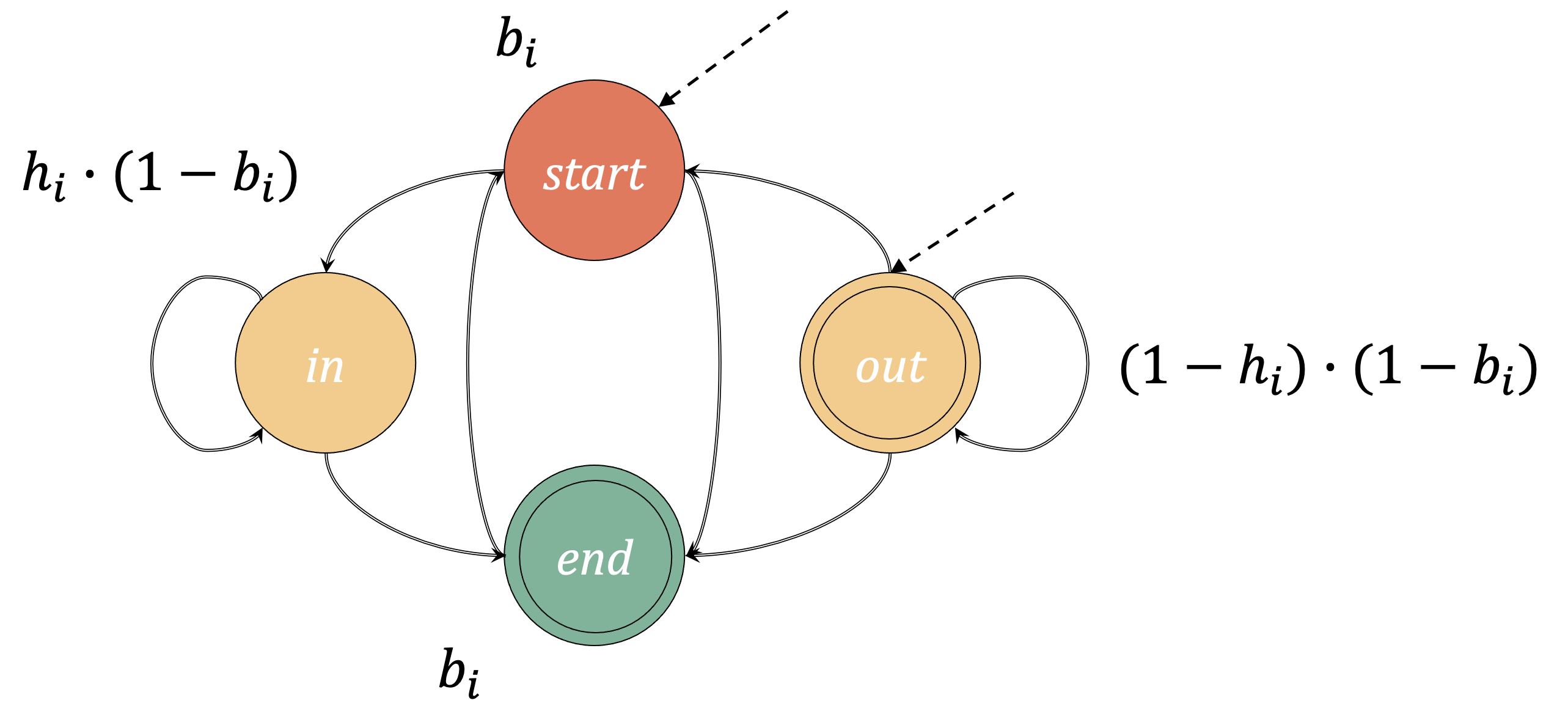}
\caption{Diagram of non-deterministic finite automaton equivalent to our proposed dynamic-programming mechanism.}
\label{fig:machine}
\end{figure}

\section{Further Explanations and Details}
\subsection{Detailed Statistics about \textit{AntHighlight} Dataset}
In this section, we will display more details about the \textit{AntHighlight} dataset in the following histograms.

\begin{figure}
\includegraphics[width=\linewidth]{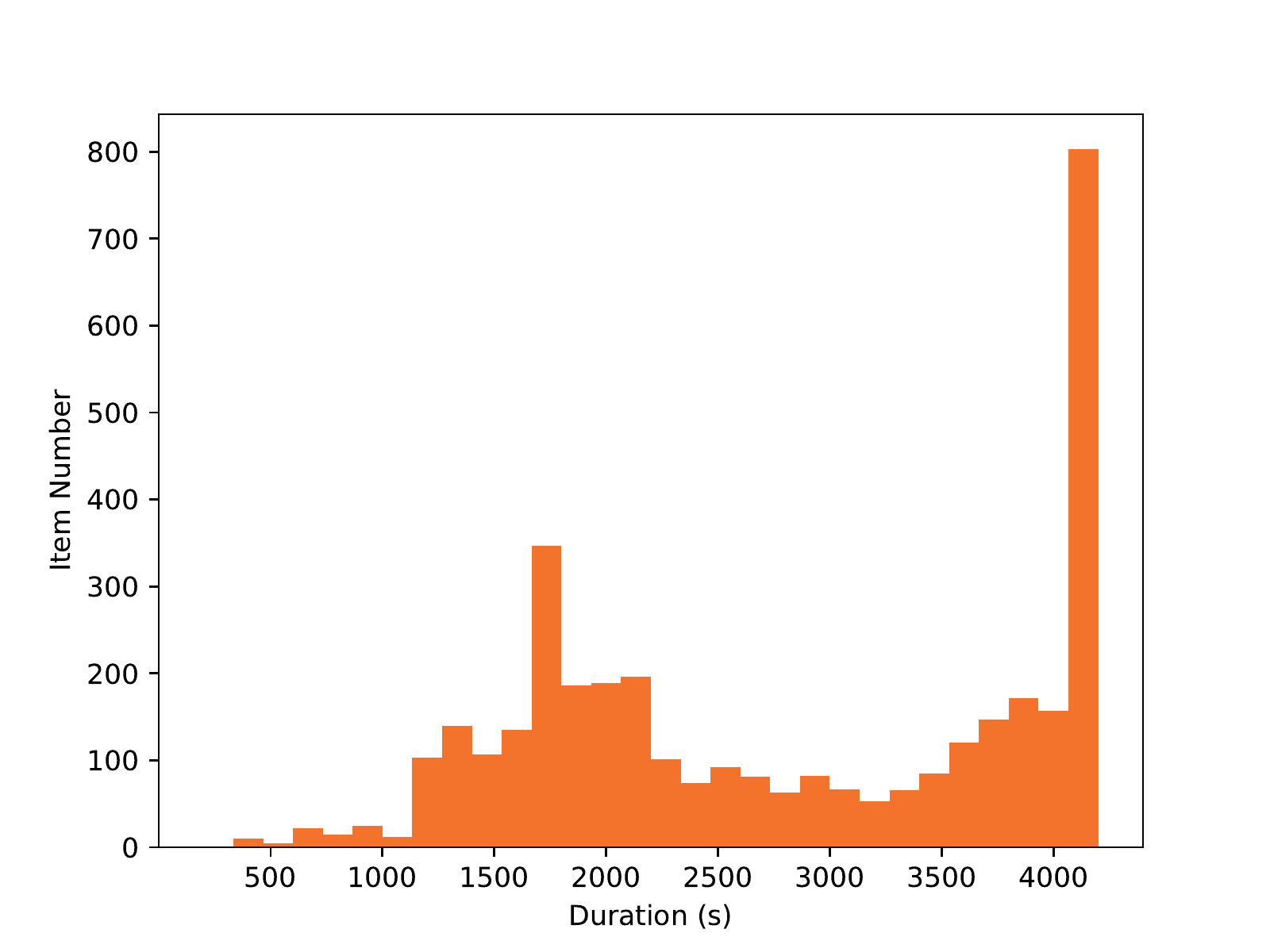}
\caption{Histogram of record duration.}
\end{figure}

\begin{figure}
\includegraphics[width=\linewidth]{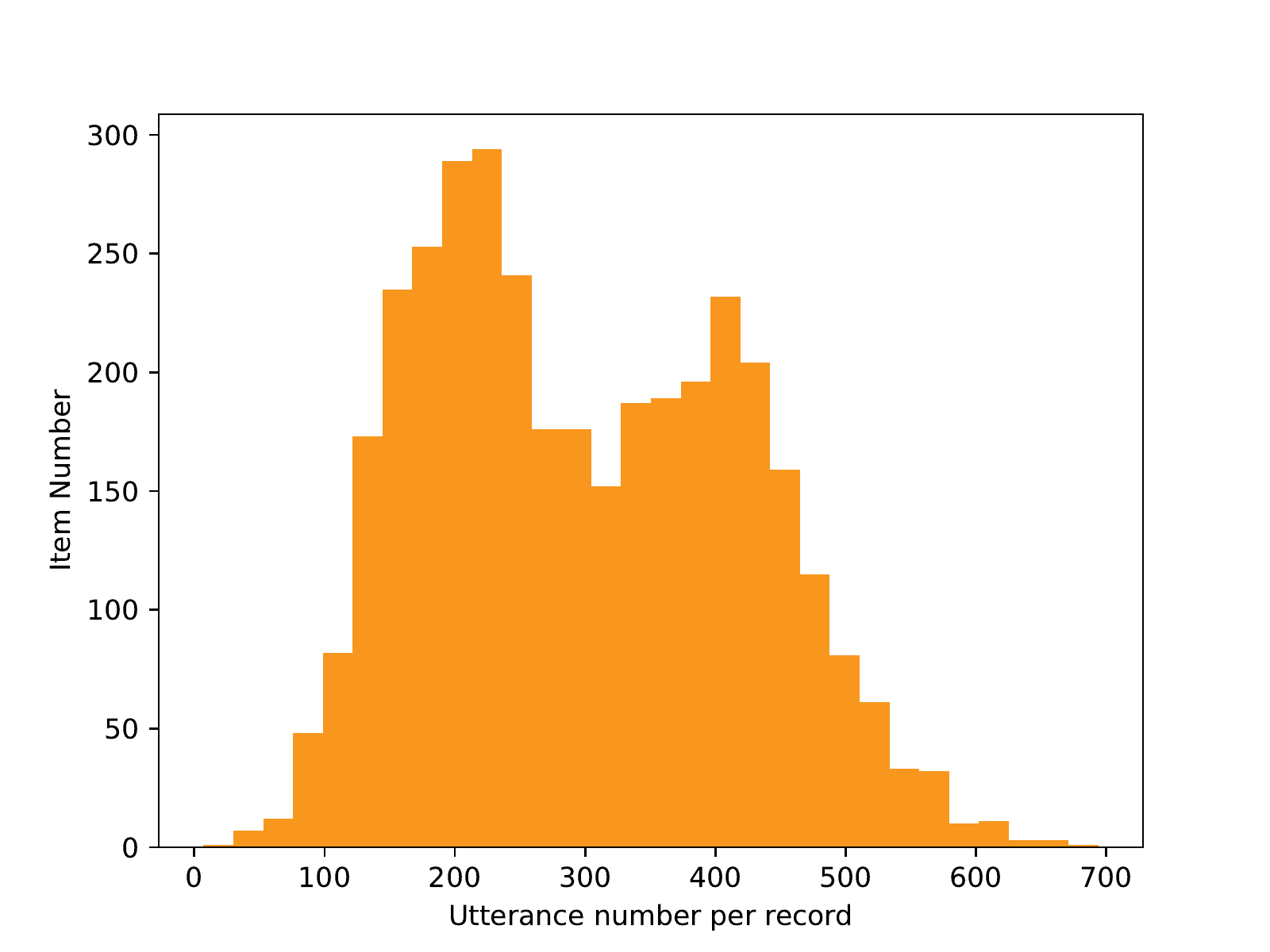}
\caption{Histogram of utterance number per record.}
\end{figure}

\begin{figure}
\includegraphics[width=\linewidth]{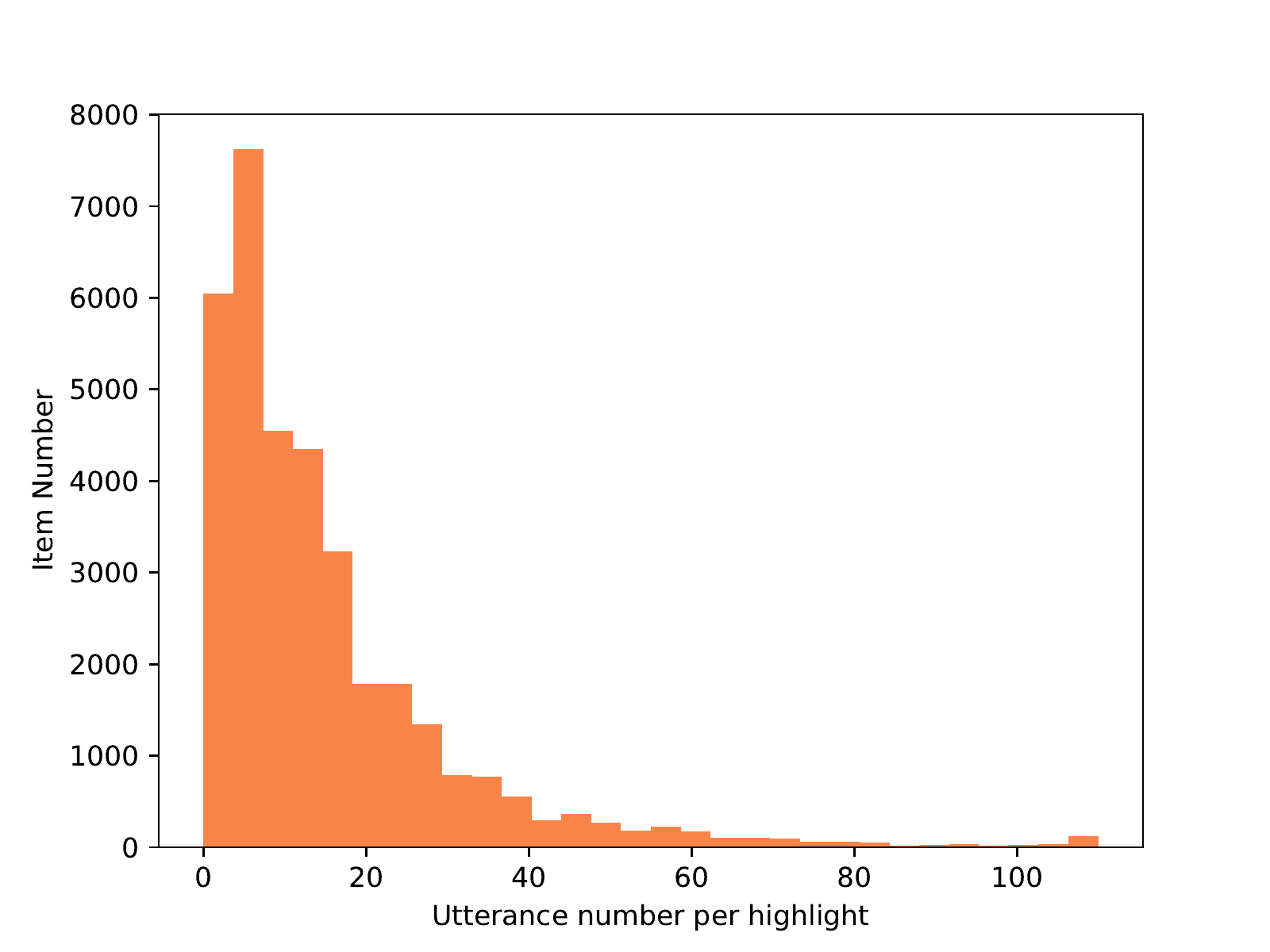}
\caption{Histogram of utterance number per highlight.}
\end{figure}

\begin{figure}
\includegraphics[width=\linewidth]{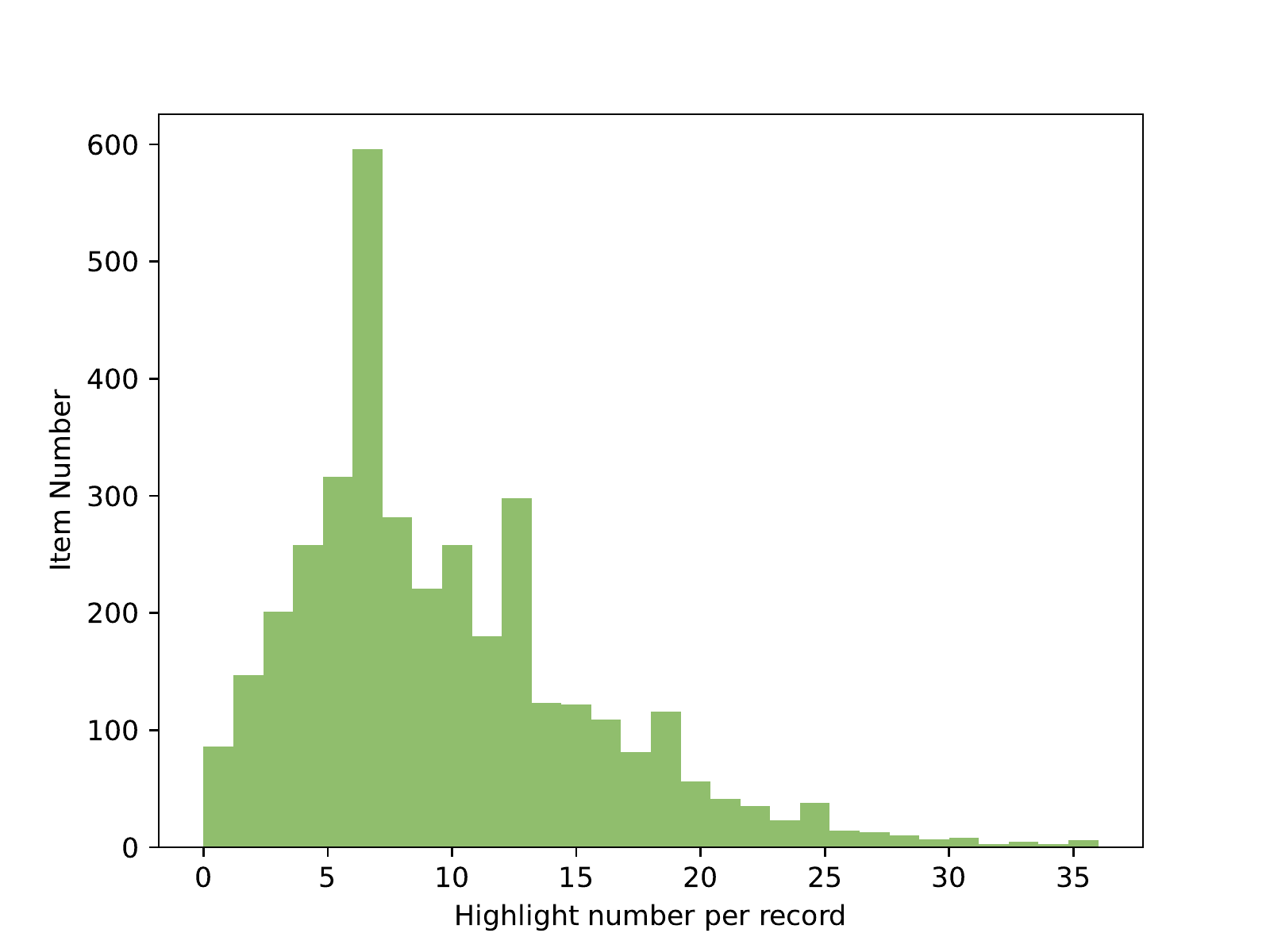}
\caption{Histogram of highlight number per record.}
\end{figure}

\begin{figure}
\includegraphics[width=\linewidth]{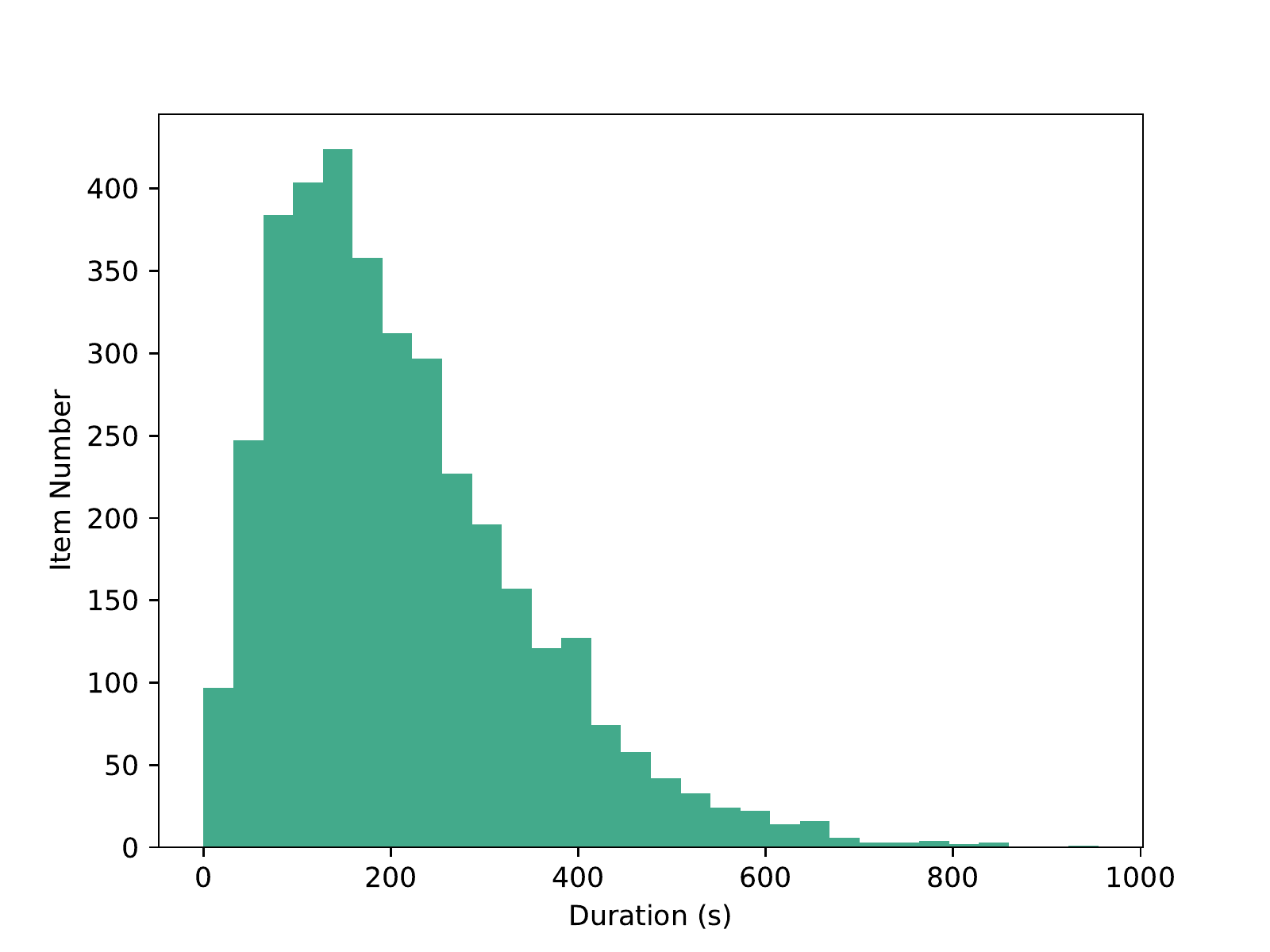}
\caption{Histogram of highlight duration.}
\end{figure}

\begin{figure}
\includegraphics[width=\linewidth]{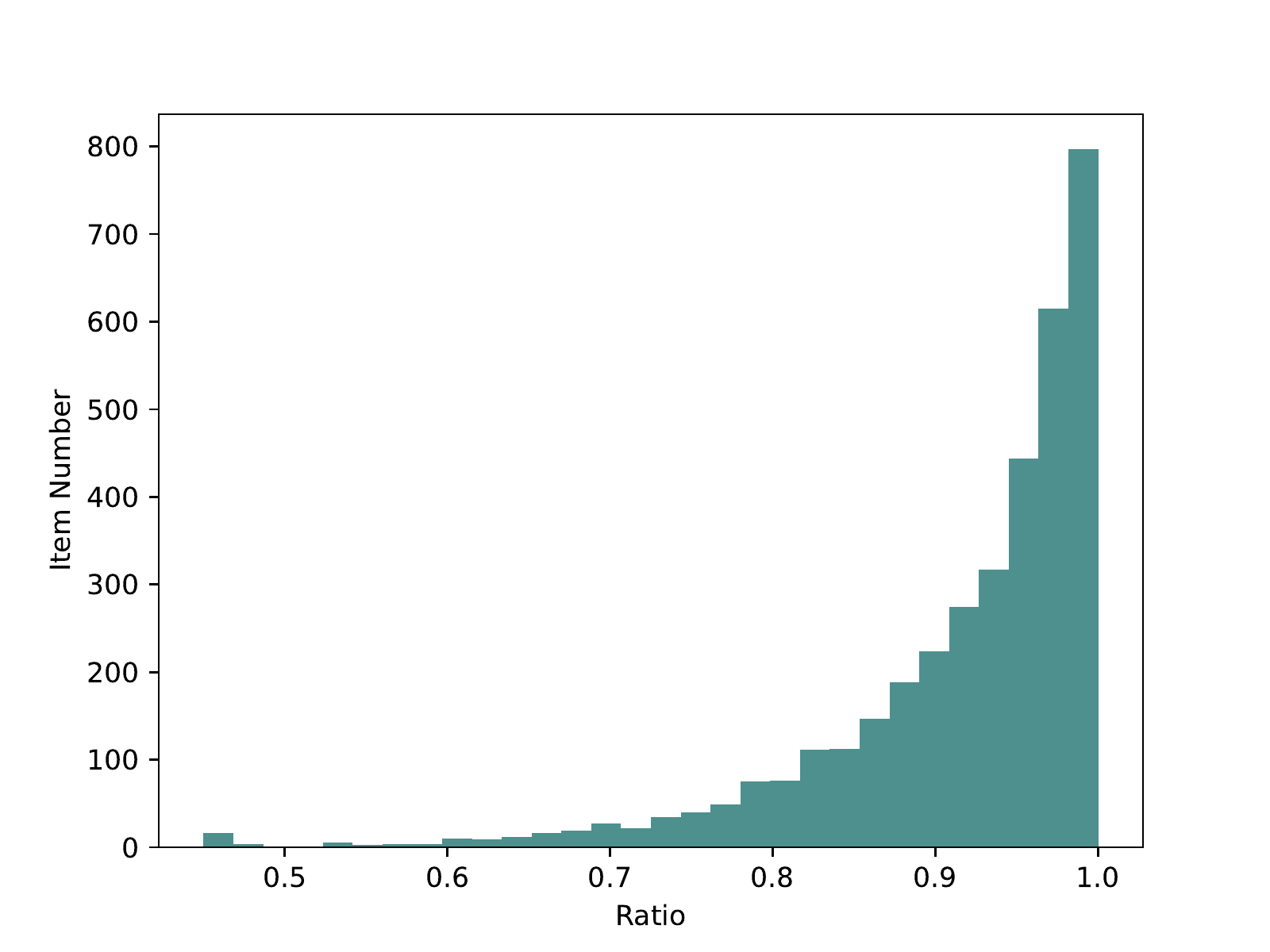}
\caption{Histogram of utterance ratio per record.}
\end{figure}

\begin{figure}
\includegraphics[width=\linewidth]{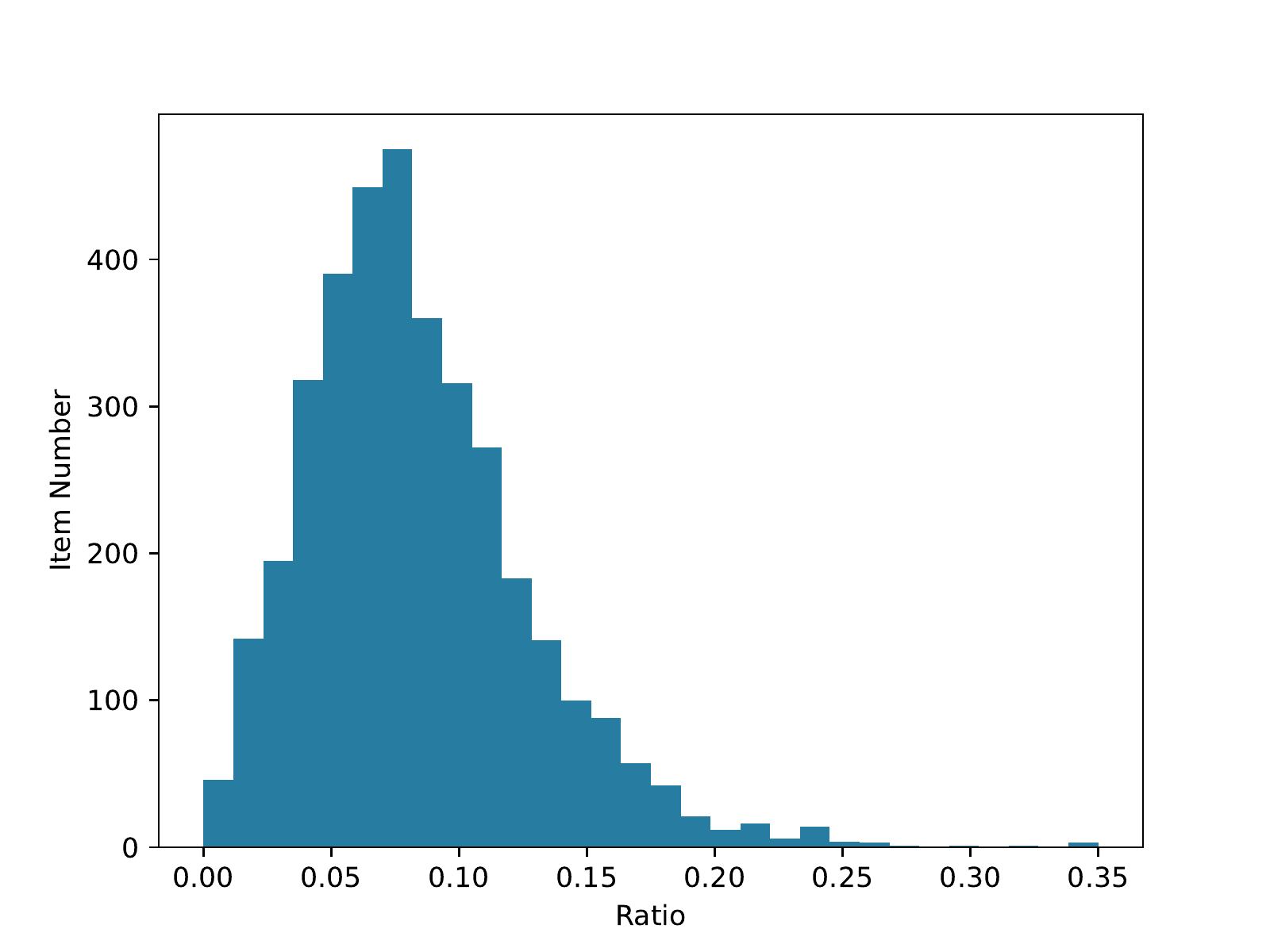}
\caption{Histogram of highlight ratio per record.}
\end{figure}

\bibliographystyle{named}
\bibliography{ijcai22}